\documentclass{aa}
\usepackage[varg]{txfonts}

\usepackage{epsfig,graphicx,natbib,url,twoopt}
\usepackage{hyperref}          
\hypersetup{
 colorlinks=true,  
 urlcolor=blue,    
 linkcolor=red,     
 citecolor=blue 
}
\usepackage{float}
\setcitestyle{citesep={,}} 

\def\deg{\hbox{$^\circ$}}      

\def\Halpha{\mbox{H\hspace{0.1ex}$\alpha$}}
\def\Hbeta{\mbox{H\hspace{0.1ex}$\beta$}}

\def\MgK{\ion{Mg}{ii}~k}
\def\MgH{\ion{Mg}{ii}~h}
\def\Mgtriplet{\ion{Mg}{ii}~triplet}
\def\Siiv{\ion{Si}{iv}}

\begin{document} 
\title{Transition region response to Quiet Sun Ellerman Bombs}

\author{Aditi Bhatnagar \inst{1,2}
\and Luc Rouppe van der Voort \inst{1,2}
\and Jayant Joshi \inst{3}}      

\institute{
  Institute of Theoretical Astrophysics,
  University of Oslo, %
  P.O. Box 1029 Blindern, N-0315 Oslo, Norway
\and
  Rosseland Centre for Solar Physics,
  University of Oslo, %
  P.O. Box 1029 Blindern, N-0315 Oslo, Norway
\and
  Indian Institute of Astrophysics, 
  II Block, Koramangala, Bengaluru 560 034, India
}


\date{submitted to A\&A Mar 22, 2024 / accepted Jun 11, 2024} 

\abstract
{ 
Quiet Sun Ellerman Bombs (QSEBs) are key indicators of small-scale photospheric magnetic reconnection events. Recent high-resolution observations have shown that they are ubiquitous and that large numbers of QSEBs can be found in the quiet Sun.}
{ 
We aim to understand the impact of QSEBs on the upper solar atmosphere by analysing their spatial and temporal relationship with the UV brightenings observed in transition region diagnostics.}
{ 
We analyse high-resolution \Hbeta\ observations from the Swedish 1-m Solar Telescope and utilise $k$-means clustering to detect 1423 QSEBs in a 51~min time series. We use coordinated and co-aligned observations from the Interface Region Imaging Spectrograph (IRIS) to search for corresponding signatures in the 1400~\AA\ slit-jaw image (SJI) channel and in the \Siiv\ 1394~\AA\ and \ion{Mg}{ii} 2798.8~\AA\ triplet spectral lines. We identify UV brightenings from SJI 1400 using a threshold of 5$\sigma$ above the median background.}
{ 
We focused on 453 long-lived QSEBs ($>1$~min) and found 67 cases of UV brightenings from SJI 1400 occurring near the QSEBs, both temporally and spatially. Temporal analysis of these events indicates that QSEBs start before UV brightenings in 57\% of cases, while UV brightenings lead in 36\% of instances. The majority of the UV brightenings occur within 1000~km from the QSEBs in the direction of the solar limb.
We also identify 21 QSEBs covered by the IRIS slit, with 4 of them showing emissions in both or one of the \Siiv\ 1394~\AA\ and \ion{Mg}{ii} 2798.8~\AA\ triplet lines, at distances within 500~km from the QSEBs in the limb direction.}
{ 
We conclude that a small fraction (15\%) of the long-lived QSEBs contribute to
localized heating observable in transition region diagnostics, indicating a minimal role in the global heating of the upper solar atmosphere.}

\maketitle

\section{Introduction}
\label{sec:introduction}
Ellerman bombs (EBs) were first identified by \citet{1917ApJ....46..298E}, who termed them as ``solar hydrogen bombs''. 
These occur at sites with opposite magnetic polarities in close proximity where magnetic reconnection takes place. 
EBs are prominent in the hydrogen Balmer-$\alpha$ line (\Halpha) line as small-scale, short-lived enhancements of the spectral line wings and are often observed in solar active regions with magnetic flux emergence 
\citep[see, e.g.,][]{2002ApJ...575..506G,
2004ApJ...614.1099P, 2006ApJ...643.1325F, 2007A&A...473..279P, 2008PASJ...60..577M, 2008ApJ...684..736W,2017A&A...598A..33L}.
When viewed towards the limb at high spatial and temporal resolution, they appear as bright, rapidly flickering flame-like structures which are rooted in the photosphere \citep{2011ApJ...736...71W, 2013JPhCS.440a2007R,2015ApJ...798...19N}. 
They have a characteristic moustache shaped spectral profile of the \Halpha\ line \citep{1964ARA&A...2..363S}, 
with enhanced emission in the wings and a dark absorption line core. 
\cite{2013ApJ...774...32V} found that EBs also show signatures in the AIA 1600 and 1700 \AA\ passbands.

\citet{2016A&A...592A.100R} discovered Quiet Sun Ellerman-like brightenings (QSEBs) through \Halpha\ observations made with the Swedish 1-m Solar Telescope \citep[SST,][]{2003SPIE.4853..341S} in regions away from active regions. 
\citet{2020A&A...641L...5J} found that the QSEBs are much more prevalant than earlier estimates using for the first time \Hbeta\ observations from SST. 
They found that around half a million QSEBs can be present on the Sun at a given time. 
In a recent study by \cite{2024A&A...683A.190R}, this estimate has been increased to the presence of 750,000 QSEBs at any given time. 
They employed the shorter wavelength H$\varepsilon$ line which provides higher spatial resolution and larger contrast leading to an increase in the number of QSEB detections.
The large number of QSEB detections raises the question about their possible impact on energy transfer in the lower solar atmosphere.
 
Ultraviolet (UV) bursts can be identified as intense brightenings in Interface Region Imaging Spectrograph (IRIS) observations of the \Siiv\ spectral lines \citep{2014Sci...346C.315P}.
They are typically less than 2\arcsec\ in size and their duration can extend from seconds to more than an hour \citep{2018SSRv..214..120Y}. 
These phenomena occur in regions of opposite magnetic polarities, analogous to the conditions of EBs. 
The spectral signatures of \Siiv\ are often characterized by broadened emission lines with weak absorption blends from neutral or singly ionised species, leading to the inference that the hot UV bursts are located below the colder chromospheric canopy of fibrils \citep{2014Sci...346C.315P}.
\cite{2015ApJ...812...11V} provided examples of co-spatial and co-temporal EBs and UV bursts using IRIS spectra, particularly in the \Mgtriplet\ and \Siiv\ lines indicating that the tops of EBs might reach transition region-like temperatures, significantly higher than previously thought.
\cite{2016ApJ...824...96T} found ten UV bursts either directly or potentially linked to EBs, characterized by notable \Halpha\ wing brightening without any signature in the line core. 
The examination of EBs by \cite{2017A&A...598A..33L} further expands our understanding of their properties, with a specific focus on neutral helium triplet lines. 
They find that EBs must reach high temperatures, estimated to be in the range of $2 \times 10^{4}$ to $10^{5}$ K. 

\cite{2019A&A...626A..33H} employed a 3D radiative magneto-hydrodynamic (MHD) Bifrost model of a network region with a strong magnetic field injected 2.5 Mm beneath the solar surface. 
Their study reveals that EBs and UV bursts can occur either simultaneously or with some time difference along extended current sheets and are interconnected within the same reconnection system, with EBs occurring from the low photosphere up to 1200~km height and UV bursts ranging from 700 km to 3 Mm above the photosphere. 
They also noted potential offsets in EB and UV burst occurrences, influenced by the orientation of the current sheet and the viewing angle.
\cite{2020A&A...633A..58O} and \cite{2019ApJ...875L..30C} provide the observational perspective of the above scenario and explore the temporal and spatial interplay between EBs and their counterparts in the chromosphere and transition region.
Until now, the only reported transition region response to a QSEB was by \cite{2017ApJ...845...16N}, based on coordinated \Halpha\ observations from the SST and IRIS.
In this work, we further explore the connection between QSEBs and higher atmospheric diagnostics. 
In particular, this study explores the spatial and temporal relationship between QSEBs and associated brightenings in the IRIS \Siiv\ observations to understand the dynamics between these events.

\section{Observations}
\label{sec:observations}

A quiet Sun region close to the North limb of the Sun was observed on 22 June 2021 by the CRisp Imaging SpectroPolarimeter (CRISP) and The CHROmospheric Imaging Spectrometer (CHROMIS) \citep{2008ApJ...689L..69S} at the Swedish 1-m Solar Telescope \citep[SST,][]{2003SPIE.4853..341S}. 
The observing angle $\mu$ was 0.48 and the heliocentric coordinates were $(x,y) = (4\arcsec, 827\arcsec)$. 
The observation started at 08:17:52 UT and ended at 09:08:32 UT, the temporal duration being 51 min. 
CHROMIS was used to obtain the \Hbeta\ spectral line at 4861~\AA\ across 27 line positions spanning $\pm$2.1~\AA\ around the line core. 
Within the inner wings at $\pm$1.0~\AA\, the sampling was done in 0.1~\AA\ increments, with intervals becoming more spaced out in the outer wings to avoid blends.
The CHROMIS observations have a temporal cadence of 7~s, pixel scale of 0\farcs038 and cover an area of 66\arcsec $\times$ 42\arcsec.
CRISP ran a program sampling the \ion{Fe}{i} 6173~\AA, and \ion{Ca}{ii} 8542~\AA\ spectral lines at a cadence of 19~s. 
\ion{Ca}{ii} 8542~\AA\ was sampled in 4 line positions without polarimetry. 
Observations of the \ion{Fe}{i} 6173~\AA\ spectral line with polarimetry covered 13 line positions, spanning from $\pm$0.16~\AA\ in 0.04~\AA\ intervals, extending to $\pm$0.24~\AA\ and $\pm$0.32~\AA\, along with the continuum located +0.68~\AA\ away from the line core. 
In each line position, eight exposures were captured for each of the four polarization states, which, when considering the continuous cycling through four different states of the liquid crystal modulators, resulted in a total of 32 exposures per line position. 
The final CHROMIS and CRISP datasets used for the analysis are obtained through the SSTRED reduction pipeline \citep[]{2015A&A...573A..40D, 2021A&A...653A..68L}.
High data quality was achieved with the aid of the SST adaptive optics system \citep{2023arXiv231113690S} 
and multi-object multi-frame blind deconvolution \citep[MOMFBD,][]{2005SoPh..228..191V} image restoration. 
For generating maps of the line-of-sight magnetic field strength ($B_\mathrm{LOS}$), Milne-Eddington inversions were performed on the \ion{Fe}{i} 6173~\AA\ observations using an inversion code by \cite{2019A&A...631A.153D}.
To determine the noise level for the line-of-sight magnetic field $B_\mathrm{LOS}$, we selected a small area away from strong magnetic field concentrations. 
The noise level was then defined as the standard deviation within this selected region: 6.4~G.
The lower-resolution CRISP data were aligned to the higher-resolution CHROMIS data by performing cross-correlation on the wideband channels. The lower cadence CRISP \ion{Fe}{I} magnetograms were matched to the 7~s cadence CHROMIS data with nearest neighbour interpolation.

The quiet Sun region was also co-observed with the Interface Region Imaging Spectrograph \citep[IRIS,][]{2014SoPh..289.2733D}. 
IRIS was running a “Large dense 4-step raster” program (OBS-ID 3633109417) and observed a bigger field of view (FOV) compared to the SST observation.
The spectrograph slit covered an area of 1\arcsec\ $\times$ 119\arcsec\ by building up a raster of 4 slit positions with 0\farcs35 separation. 
The raster cadence was 36~s with a step cadence of 9~s and each slit position has an exposure time of 8~s. 
The IRIS Slit Jaw Images (SJIs) cover a FOV of 120\arcsec\ $\times$ 119\arcsec\ and were taken with a cadence of 18~s for the 2796~\AA\ channel (covering the chromospheric \MgK\ line core) and 1400~\AA\ channel (covering the two \Siiv\ transition region lines).
The IRIS data were spatially binned and have a pixel size of 0\farcs33. 
The SJIs were first rotated to match the orientation of the SST images.
We then used the SJI 2796~\AA\ \MgK\ for alignment of IRIS data to SST datasets by blowing up the IRIS data to CHROMIS pixel scale and performed cross-correlation between the wing positions of aligned \ion{Ca}{ii}~8542~\AA\ and the SJI 2796~\AA.
The aligned IRIS data were then cropped to match the FOV of the CHROMIS data. 
The error in the co-alignment of these datasets is at least 1 IRIS pixel or better than that. 
\cite{2020A&A...641A.146R} discuss the possible factors that can induce errors in the alignment.

In addition to a detailed analysis of the 22 June 2021 data, we have also performed some analysis of a similar dataset from 25 July 2021.
This is a 40~min time sequence of a quiet Sun region at $\mu = 0.57$. 
Both CRISP, CHROMIS, and IRIS were running the same observing programs. 
The seeing quality was similar to the June data. 
This July dataset was used to verify some of the results, such as the number of QSEB and UV brightening detections and spatial offsets between them. 
Furthermore, the IRIS raster spectra were explored for the possible presence of a strong UV burst-like event. 

\section{Method of analysis} 

\label{sec:methods}
\subsection{Detection of QSEBs}

We have implemented the \textit{k}-means clustering algorithm on our \Hbeta\ data, first employed by 
\citet{2020A&A...641L...5J} 
for automatic detection of QSEBs.
\textit{k}-means clustering is a machine learning technique which groups a dataset into a predefined number of \textit{k} clusters. 
During the initialization phase, we opted for the \textit{k}-means++ method \citep{Arthur2007}. 
More details on this are available in \cite{2022A&A...664A..72J}. 
After establishing initial clusters, each cluster is represented by its cluster center (representative profile), which is the mean of all data points assigned to that cluster. 

Before performing the clustering of the full dataset, we created a subset of data to make a cluster model that was used to predict the clustering of all 426 time steps. 
We started the clustering with 100 clusters on this subset as in \citep{2022A&A...664A..72J}, but this did not give any clusters with clear QSEB-type spectral profiles.
Therefore, we created a subset from selected profiles from the ten best-seeing scans and from frames 0--180. 
Hence, we opted for creating a biased training dataset, making sure that it contains spectral profiles representing QSEBs in the following way. 
From the chosen subset, we pick data points that show intensity enhancements in a few wavelength positions in the wings of \Hbeta\ line profile - comprising EBs and magnetic bright points. 
We also keep some profiles corresponding to 
the bright centers of granules. 
In this way, we have a dataset that for 81\% consists of profiles with strong wing intensity, and 19\% of other types of profiles.
We used this dataset to train the \textit{k}-means model and found that we could reduce the number of clusters to 64.
We identified 27 clusters that have representative profiles (RPs) similar to QSEBs.
From the selected RPs, we predict the QSEBs from the full time series. 
After this we track the QSEBs temporally, utilizing the Trackpy library of Python, which initially identifies the central coordinate of each QSEB and assigns them a unique event ID number. 
Trackpy links the events over time and records information about each event ID -- including the coordinates of the centers at each timestep, and the start and end times of the events.
Following this, we implement connected component analysis from OpenCV. 
For each event ID, a region encompassing 26 pixels in space and time \citep[as used by][]{2022A&A...664A..72J} is analysed to identify all pixels that are occupied by the QSEB. 
This analysis checks for any pixel connected to the central coordinate pixel, whether horizontally, vertically, or diagonally. 
Through this temporal and spatial tracking of QSEBs, each QSEB has an event ID number, and information about the duration and region occupied by it at each timestep, which can be used for further in-depth analysis.
To refine our event selection, we have set specific criteria: events that occur just for one timestep and have a maximum area of less than 5 pixels are excluded. 
So, any large event occurring for one timestep or any small event occurring for more than one timestep are considered as QSEBs.
Furthermore, two events are considered as a single event if they are separated by a temporal gap of no more than 35~s, which is equal to 5 timesteps. 
This criterion was decided by checking how the mask from \textit{k}-means detection evolves for a few QSEBs. Five timesteps allows for some variations in the \textit{k}-means detection. Sometimes a QSEB can go through a phase when the spectral profiles do not pass the thresholds of the detection method. This can be due to for example seeing variations. By considering 5 timesteps we are linking the same QSEB in time, and it is not too long that we connect different events in time.
Through this process, we have identified a total of 1423 QSEB events. 
We focused on longer-lived QSEBs, from the 1423 events, that have durations exceeding 8 timesteps ($\approx$1~min), resulting in a total of 453 such events. 
This is done because of the higher cadence of \Hbeta\ data and lower cadence of SJI data, giving enough SJI frames to check the evolution of UV brightening with the QSEB.

\subsection{Detection of UV brightenings}

The SJI 1400 images are full with short-lived and compact brightenings, most of them originate from acoustic waves and are related to the chromospheric bright grains observed in the \ion{Ca}{ii}~H and K lines 
\citep{1997ApJ...481..500C}. 
The signal in these grains in SJI 1400 comes mostly from the continuum rather than the \ion{Si}{iv} lines 
\citep{2015ApJ...803...44M}. 
We are interested in investigating the connection between QSEBs and the brightest SJI 1400 events that are not related to acoustic waves.
Following the approach of \cite{2018SSRv..214..120Y}, who argued against a universal threshold for event identification in favor of dataset-specific criteria, we have employed a threshold of 5$\sigma$ above the median background to identify significant \Siiv\ emissions in the SJI 1400, henceforth referred to as ``UV brightenings''.
We found that a value of 5$\sigma$ was most effective in identifying brightenings related to QSEBs. 
The threshold of 5$\sigma$ is a conservative criterion to consider only the strongest events and avoid any ambiguous connections between the QSEBs and UV brightenings. It is possible that there could be more UV brightenings associated with the QSEBs by keeping a lower threshold, but these would not be the strongest events.
We track and label these UV brightenings in a similar way as we did for the QSEBs. 
This approach resulted in the detection of 1978 such events. 
From the tracked QSEBs and UV brightening events, we identify those instances where UV brightenings are in immediate proximity to the 453 long-lived QSEBs, confined to a $4\arcsec \times 4\arcsec$ area around the long-lived QSEB, and occur within 35~s of the start or end of the long-lived QSEB.
After applying the above conditions, 199 of the 453 QSEBs have a nearby UV brightening. Out of the 1978 5$\sigma$ UV brightenings, 272 are associated with the 199 QSEBs. 

While most of the SJI 1400 signal in the chromospheric bright grains comes from the continuum, 
\citet{2015ApJ...803...44M} found that sometimes grains have shock signatures in the \Siiv\ lines. 
To eliminate these short-lived shock-related grains, we employed a Fourier filter on the SJI 1400 dataset, adopting a 5 km~s$^{-1}$ cutoff as utilized by \cite{2018ApJ...857...48G} in their examination of quiet Sun atmosphere heating through internetwork magnetic field cancellations.
Due to the presence of noise in the SJI data, the Fourier filtering creates artefacts around sharp peaks from cosmic ray hits to the detector. 
These artefacts influence the intensity threshold for determining the UV brightenings.
Hence, instead of using the filtered SJI data, we check for those 272 UV brightening events to see if there is a considerable change in the intensity of the tracked events in the original data and in the filtered data and remove those events.
This reduces the number of UV brightenings to 233, with 159 QSEBs associated with them.
By applying the above criteria, we 
significantly reduced the risk of 
chance alignment of a shock with a QSEB. Also, by considering QSEBs with durations of more than 1~min we only focus on the strongest events, for which we can clearly see the evolution of the UV brightenings with the QSEB. 
We also removed many events for which the connection between QSEB and UV brightening was unclear. These are mentioned in the following paragraph. The removed QSEBs and UV brightenings could also be related, but we only focus on clear associations between these events for analysis.

In the analysis of the 159 QSEB - 233 UV brightening events, we identified instances where the association between QSEBs and UV brightenings was ambiguous, as some of the UV brightenings could not be associated with just the QSEB. 
For example, QSEBs amidst dense concentrations of magnetic bright points and large extended areas with \Siiv\ emission. 
Other examples are isolated events where QSEB profiles are within an area with magnetic bright point profiles, and QSEBs adjacent to spicules not originating at the site of the QSEB. 
All these cases also have UV brightenings near them, which makes it difficult to associate them unambiguously with QSEBs.
These observations led us to exclude such events, focusing on pairs where the UV brightening could be directly linked with the QSEB. 
Note here that we have included those events where the spicules originate very close to the QSEBs.
This resulted in a total of 67 QSEB-UV brightening event pairs, which we use to study their spatial and temporal relationship. 

Additionally, we focused on instances where the IRIS slit was positioned directly over QSEBs, finding 21 occurrences of this scenario. 
For these cases, we analyzed the emission characteristics in the \Siiv\ 1394~\AA\ and \ion{Mg}{ii} 2798.8~\AA\ triplet spectral lines. 
Our analysis led to the identification of 4 QSEBs that exhibited notable enhancements in either or both \Siiv\ 1394~\AA\ and \ion{Mg}{ii} 2798.8~\AA\ triplet lines.

\begin{figure*}
    \centering
    \includegraphics[width=\linewidth]{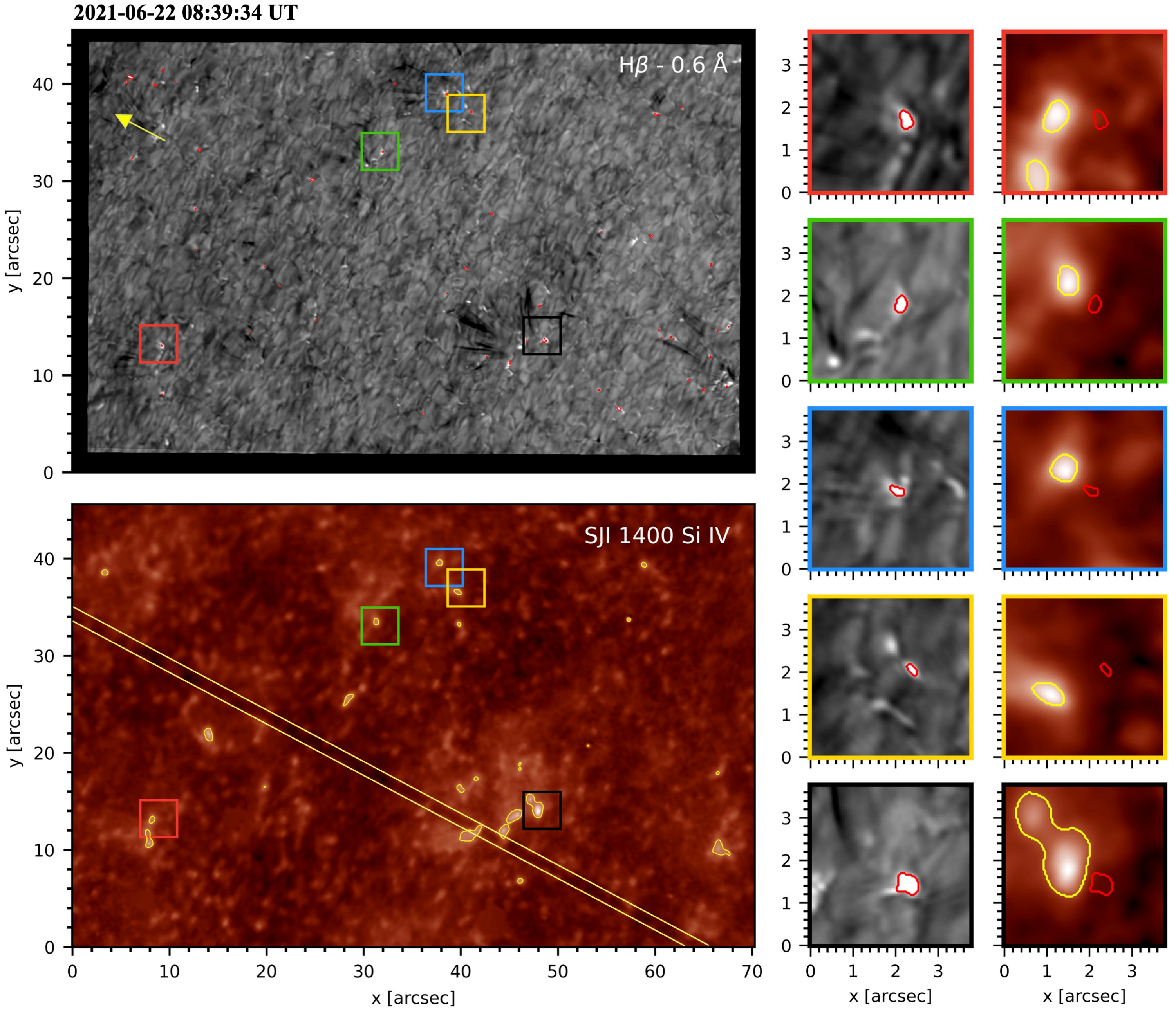}
    \caption{Examples of QSEBs in the \Hbeta\ wing and their associated UV brightenings in the aligned SJI 1400. The QSEBs are outlined with red contours in the top left panel. The $5\sigma$ brightenings are marked with yellow contours in the SJI 1400 bottom left panel. The colored boxes in the left panels are shown at higher magnification in the right columns, which display both the QSEBs and the UV brightenings. The zoomed-in boxes of SJI 1400 also show the red contours of the QSEBs near the UV brightenings. The yellow diagonal lines over the SJI mark the area that is covered by the IRIS spectrograph raster. 
    The direction of the solar limb is shown by the yellow arrow in the \Hbeta\ wing image.}
    \label{fig:1}
\end{figure*}
\begin{figure*}
    \centering
    \includegraphics[width=\linewidth]{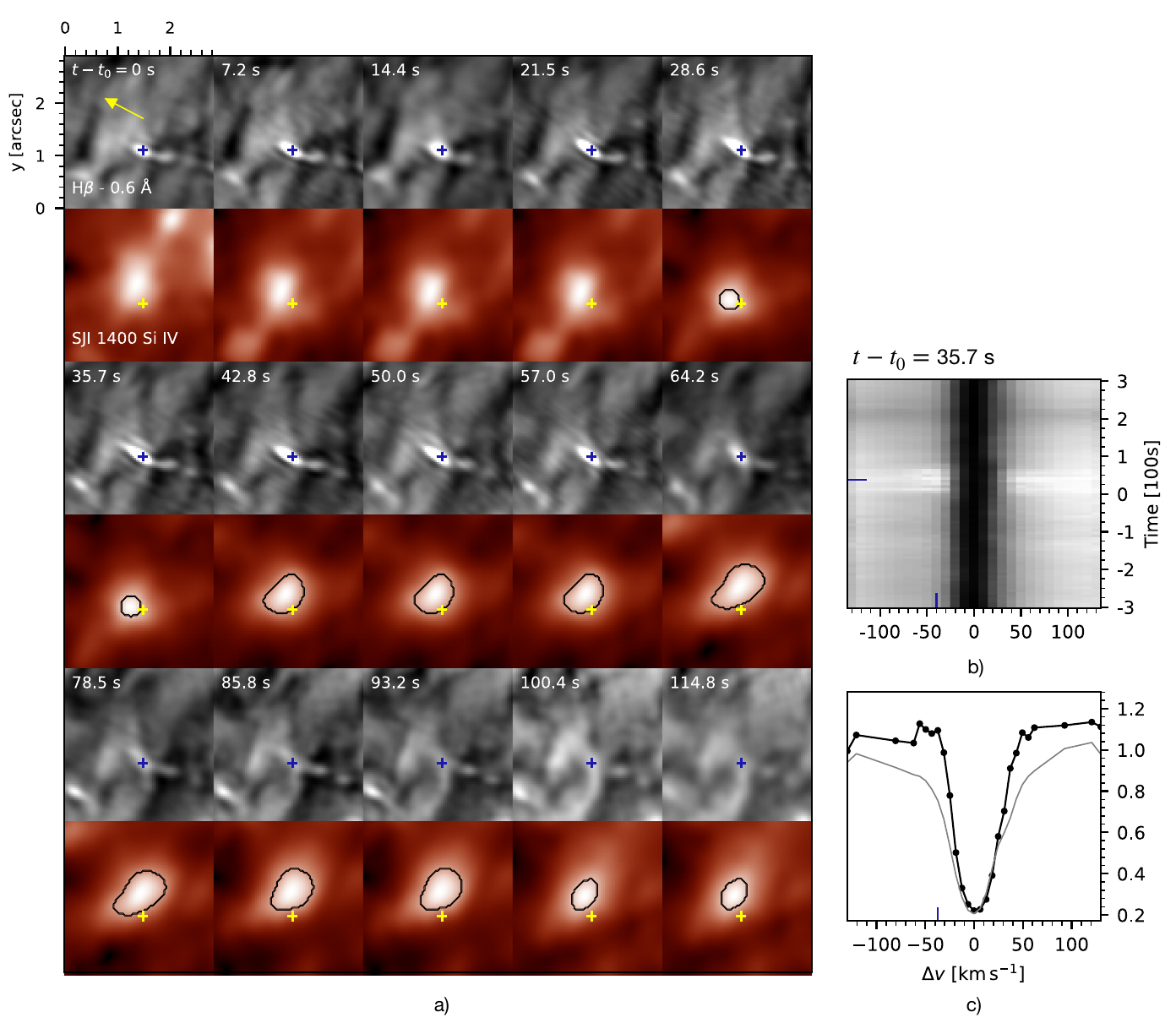}
    \caption{Temporal evolution of a QSEB and associated UV brightening. The dark blue plus symbol marks the position of the QSEB in the \Hbeta\ $-$0.6~\AA\ wing and a yellow plus symbol marks its position in the SJI 1400. Black contours in the SJI 1400 denote the UV brightening. The $\lambda t$ diagram to the right in panel b) shows the spectral evolution at the pixel location marked with a plus in the \Hbeta\ wing. The light gray profile is an averaged reference profile. The vertical blue markers in panels b) and c) denote the \Hbeta\ line wing position for which the QSEB is shown in a). An animation of this figure, which shows the evolution of the QSEB with the UV brightening along with the \Hbeta\ spectral profile, is available in the online material (see \url{http://tsih3.uio.no/lapalma/subl/qseb_uv/movies/bhatnagar_fig2_movie.mp4}).}
    \label{fig:2}
\end{figure*}

\begin{figure}
    \centering
    \includegraphics[width=\linewidth]{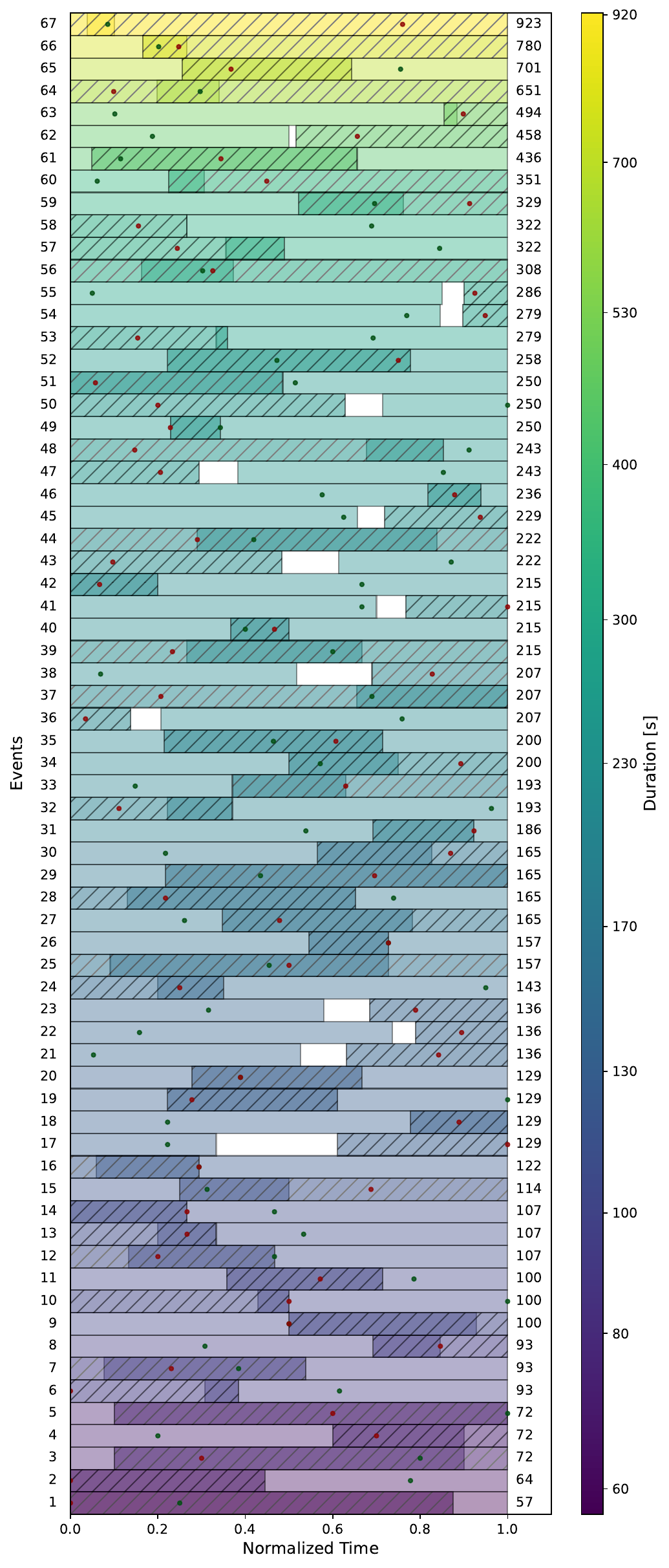}
    \caption{Temporal evolution of the 67 QSEBs and UV brightening pairs. The events are shown in order of increasing duration. The combined duration of each event pair is given in seconds to the right of each bar. The duration of each QSEB and UV brightening is normalised to a common bar length. For each case, the lighter region of the bar denotes the time when the QSEB occurs, and the region marked with diagonal lines denotes the time when the UV brightening occurs. The darker part of the bar is the time when the QSEB and UV brightening occur together. The time of peak intensity of the QSEB is marked with a green dot, and the time of peak intensity of the UV brightening is marked with a red dot.  
    }
    \label{fig:3}
\end{figure}
\begin{figure*}
    \centering
    \includegraphics[width=0.8\linewidth]{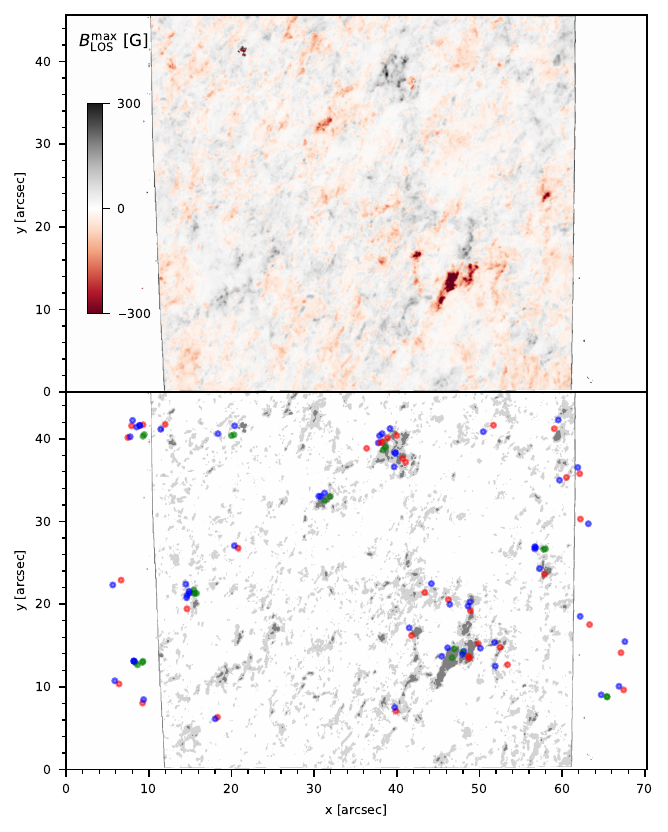}
    \caption{Spatial distribution of the QSEBs, UV brightenings and their magnetic environment. The top panel shows the extremum of $B_\mathrm{LOS}$ at each pixel for the full 51 min duration of the time series. The bottom panel shows the QSEBs (red), and the UV brightenings (blue). The green circles represent the QSEBs which have multiple occurrences during the entire duration. In the background, dark gray areas indicates regions with $|B^\mathrm{max}_\mathrm{LOS}| > 100\,\text{G}$, while light gray areas indicate regions where $50 < |B^\mathrm{max}_\mathrm{LOS}| < 100\,\text{G}$. The CRISP FOV of the \ion{Fe}{i}~6173~\AA\ data that provided the $B_\mathrm{LOS}$ data is narrower than the CHROMIS FOV.}
    \label{fig:4}
\end{figure*}

\begin{figure*}
    \centering
    \includegraphics[width=\linewidth]{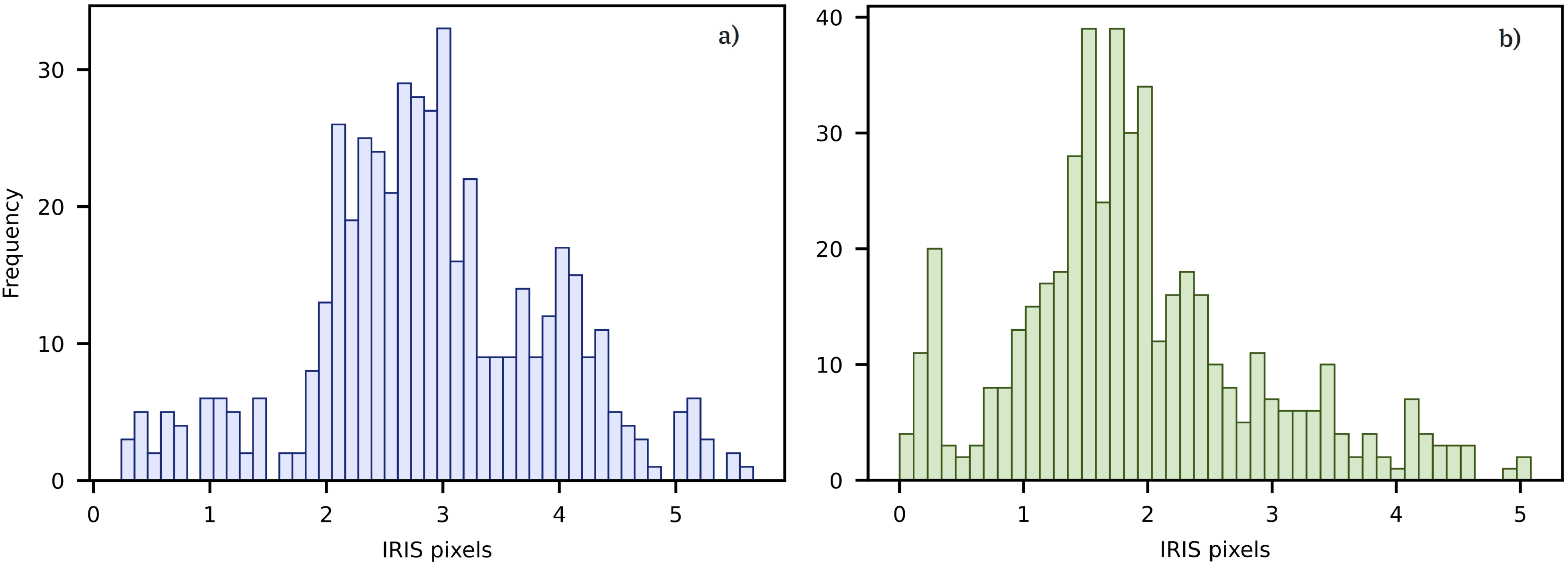}
    \caption{Histograms of the distance from the QSEB centroid to the UV brightenings. Panel a) shows the distance to the centroid of the UV brightenings, while panel b) shows the distance to the nearest boundary of the UV brightenings. The size of each bin corresponds to the angular size of 1 SST pixel (0\farcs038) and the $x$ - axis has units in IRIS pixels (1 pixel = 0\farcs33).}
    \label{fig:5}
\end{figure*}

\begin{figure*}
    \centering
    \includegraphics[width=\linewidth]{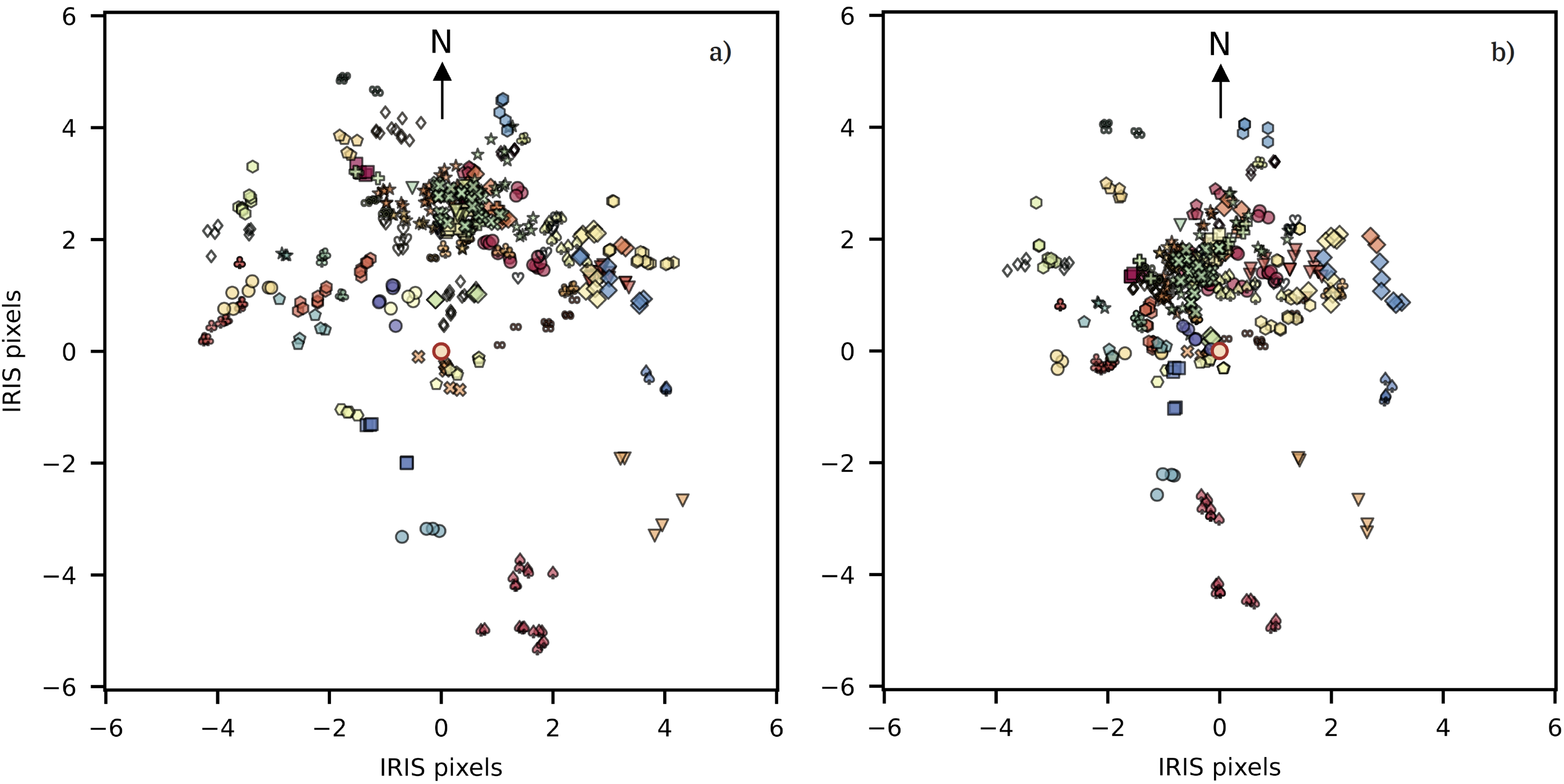}
    \caption{Spatial distribution of the 53 co-temporal UV brightenings around the QSEBs. Each QSEB is put at the center of the figure denoted by the shaded red circle. For each UV brightening, the colored markers are shown for all the overlapping instances of occurrence, and at distances and angles calculated with their respective QSEBs. Panel a) distributes the UV brightenings based on the distance of their centroids from the QSEBs. Panel b) distributes the UV brightenings based on the distance of their nearest boundaries from the QSEBs. The direction of the nearest solar limb is towards North. Here, the $x$ and $y$ axis have units in IRIS pixels (1 pixel = 0\farcs33).}
    \label{fig:6}
\end{figure*}

\begin{figure*}
    \centering
    \includegraphics[width=0.99\textwidth]{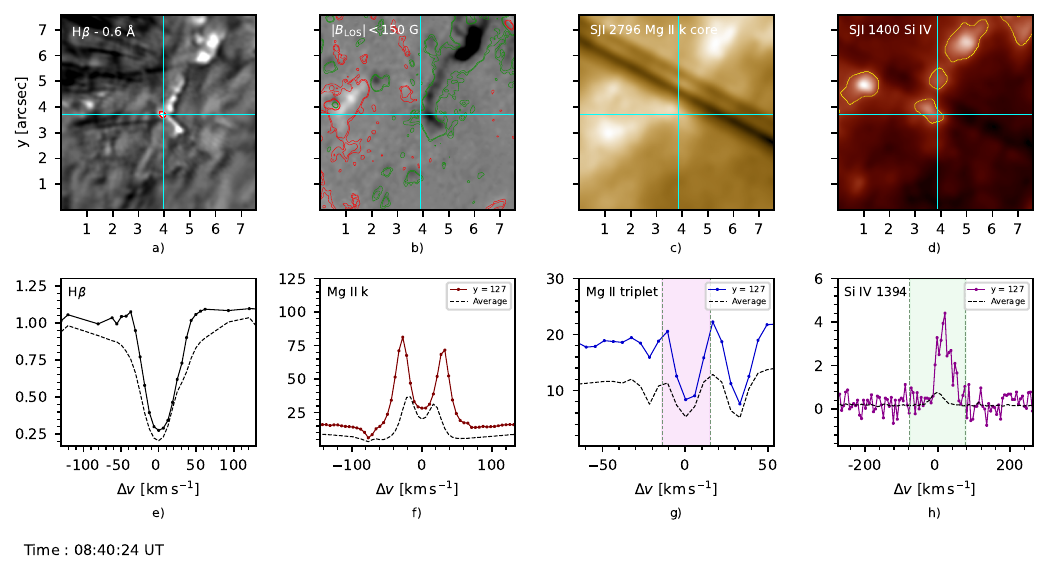}
    \caption{Snapshot of images and spectra of QSEB - 1 at the instance of maximum \Siiv\ 1394~\AA\ emission. The QSEB is located at y = 126 (in IRIS pixels) and at t = 08:40:24 UT. The top row shows the QSEB in \Hbeta\ $-0.6$~\AA, the $B_\mathrm{LOS}$ map with contours at 3$\sigma$ above the noise level, and SJI 2796 and 1400. The 5$\sigma$ contours are shown in yellow in SJI 1400. The lower row shows the normalized intensity in \Hbeta\ at the point of crossing of the cyan lines in the above \Hbeta\ $-0.6$~\AA\ image. The spectral profiles of \MgK\ 2796.39~\AA, \Mgtriplet\ 2798.8~\AA\ and \Siiv\ 1394~\AA\ are shown for the region covered by the slit at y = 127 which is at a spatial offset of one pixel from the location of QSEB and have units as DN~s$^{-1}$. An animation of this figure is available in the online material (see \url{http://tsih3.uio.no/lapalma/subl/qseb_uv/movies/bhatnagar_fig7_movie.mp4}).}
    \label{fig:7}
\end{figure*}

\begin{figure*}
    \centering
    \includegraphics{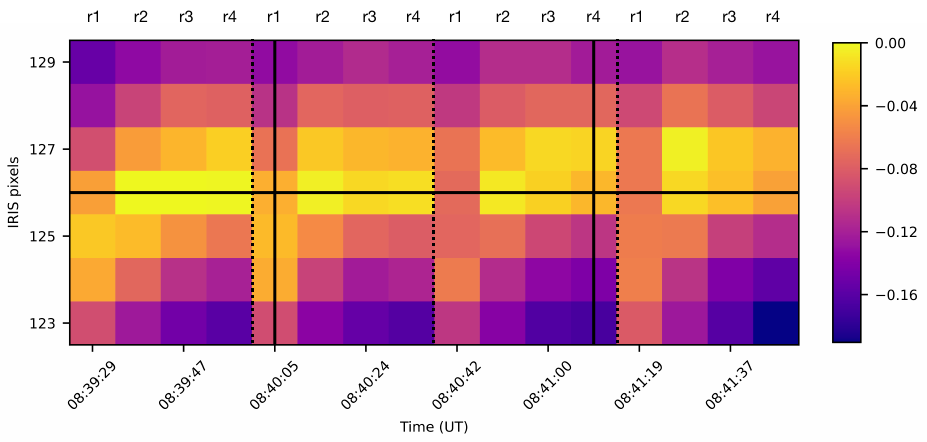}
    \caption{Time series of \ion{Mg}{ii} triplet spectroheliograms for QSEB - 1. The log of the normalised integrated line intensity for \ion{Mg}{ii}~triplet 2798.8~\AA\ line shown by the pink shaded region in Fig.~\ref{fig:7}g is displayed in the spectroheliograms. A sequence of four rasters is shown, each raster consists of four slit positions marked with r1--r4 at the top. The vertical dotted lines mark the end of each raster. The black vertical lines denote the start and end time of the QSEB and the black horizontal line denotes the location of the QSEB-1.
    }
    \label{fig:8}
\end{figure*}

\begin{figure*}
    \centering
    \includegraphics{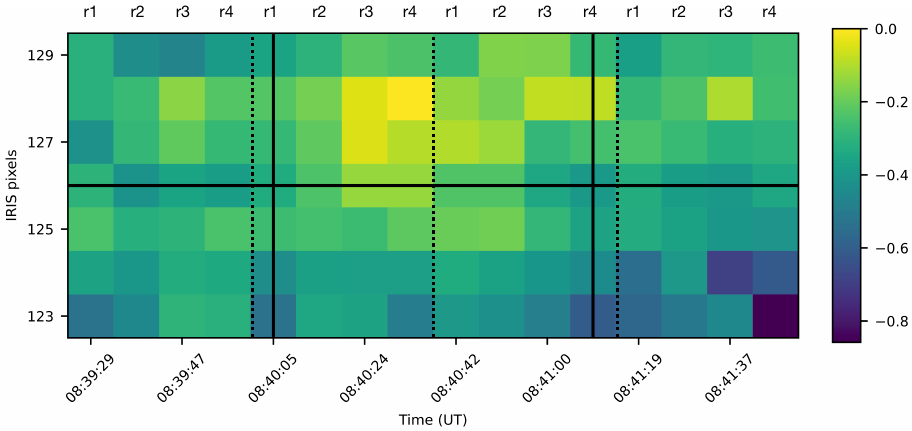}
    \caption{Time series of \ion{Si}{iv} spectroheliograms for QSEB - 1. The log of the normalised integrated line intensity for \ion{Si}{iv} 1394~\AA\ line shown by the green shaded region in Fig.~\ref{fig:7}h. Like in Fig~\ref{fig:8}, the vertical dashed lines mark the end of each of the rasters and the black vertical lines denote the start and end time of the QSEB. The black horizontal line denotes the location of the QSEB-1. 
    }
    \label{fig:9}
\end{figure*}

\begin{figure*}
    \centering
    \includegraphics[width=\textwidth]{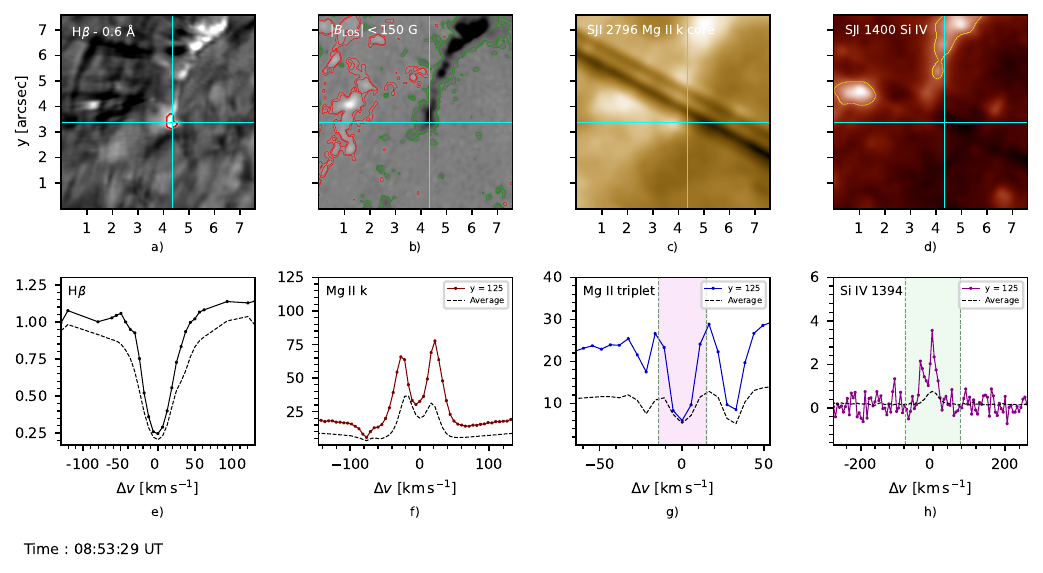}
    \caption{Snapshot of images and spectra of QSEB - 2 at the instance of maximum \Siiv\ 1394~\AA\ emission with the same format as Fig.~\ref{fig:7}. The QSEB is located at y = 124 (in IRIS pixels) at t = 08:53:29 UT. The spectral profiles of \MgK\ 2796.39~\AA, \Mgtriplet\ 2798.8~\AA, and \Siiv\ 1394~\AA\ are shown for the region covered by the slit at y = 125 which is at a spatial offset of one pixel from the location of QSEB and have units as DN~s$^{-1}$. An animation of this figure is available in the online material (see \url{http://tsih3.uio.no/lapalma/subl/qseb_uv/movies/bhatnagar_fig10_movie.mp4}).}
    \label{fig:10}
\end{figure*}

\begin{figure*}
    \centering
    \includegraphics{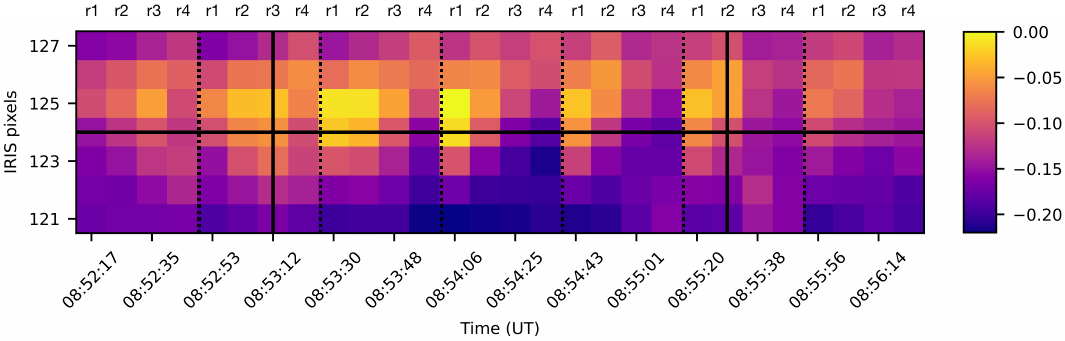}
    \caption{Time series of spectroheliograms for QSEB - 2 of the log of the normalised integrated line intensity for \Mgtriplet\ line. The black vertical lines denote the start and end time of the QSEB and the black horizontal line denotes the location of the QSEB-2.}
    \label{fig:11}
\end{figure*}
\begin{figure*}
    \centering
    \includegraphics{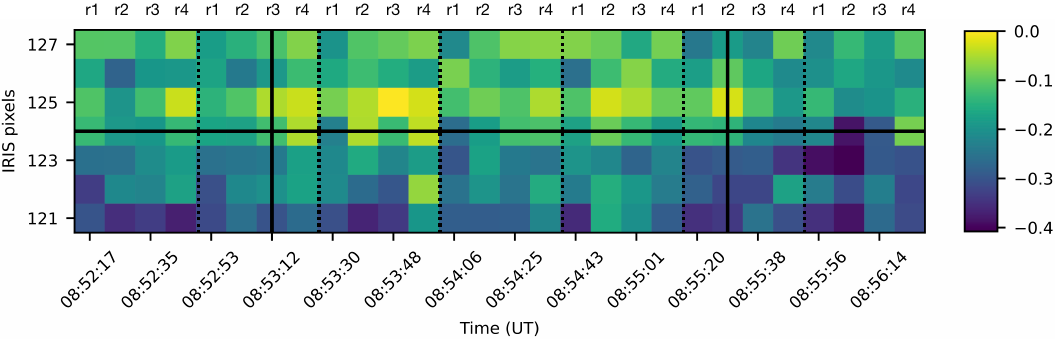}
    \caption{Time series of spectroheliograms for QSEB - 2 of the log of the normalised integrated line intensity for \Siiv\ 1394~\AA\ line. The black vertical lines denote the start and end time of the QSEB and the black horizontal line denotes the location of the QSEB-2.}
    \label{fig:12}
\end{figure*}
\begin{figure*}
    \centering
    \includegraphics{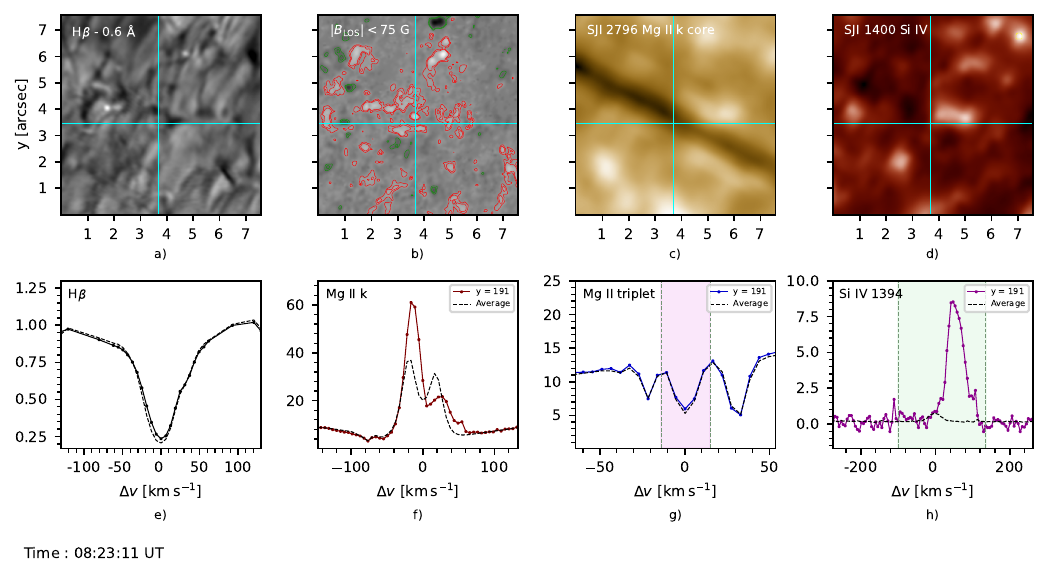}
    \caption{Snapshot of images and spectra of QSEB - 3 at the instance of maximum \Siiv\ 1394~\AA\ emission with the same format as Fig.~\ref{fig:7}. The QSEB occurs later in time and is located at y = 191 (in IRIS pixels) at t = 08:24:21 UT. The spectral profiles of \MgK\ 2796.39~\AA, \Mgtriplet\ 2798.8~\AA, and \Siiv\ 1394~\AA\ are shown for the region covered by the slit at the site of QSEB and have units as DN~s$^{-1}$. An animation of this figure is available in the online material (see \url{http://tsih3.uio.no/lapalma/subl/qseb_uv/movies/bhatnagar_fig13_movie.mp4}).}
    \label{fig:13}
\end{figure*}

\begin{figure*}
    \centering
    \includegraphics{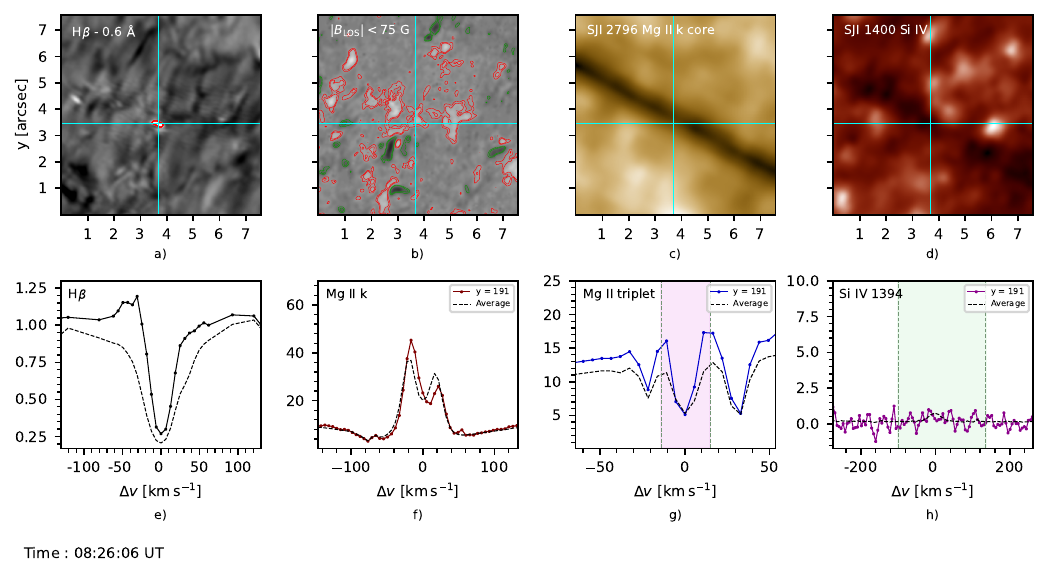}
    \caption{Snapshot of images and spectra of QSEB - 3 at the instance of maximum emission in the wings of \Hbeta\ line with the same format as Fig.~\ref{fig:7}. The QSEB is located at y = 191 (in IRIS pixels) at t = 08:26:06 UT. Panel e) shows the normalized intensity of the \Hbeta\ line. The spectral profiles of \MgK\ 2796.39~\AA, \Mgtriplet\ 2798.8~\AA\ and \Siiv\ 1394~\AA\ are shown for the region covered by the slit at the site of QSEB and have units as DN~s~$^{-1}$.}
    \label{fig:14}
\end{figure*}

\begin{figure*}
    \centering
    \includegraphics{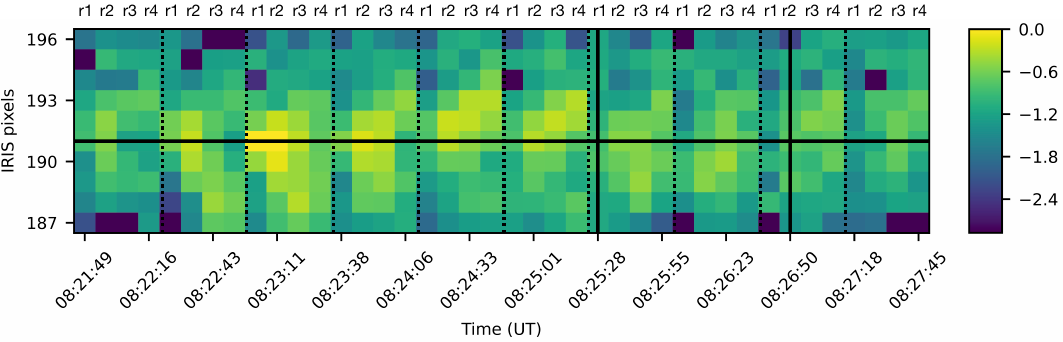}
    \caption{Time series of spectroheliograms for QSEB - 3. The log of the normalised integrated line intensity for \Siiv\ 1394~\AA\ line is considered over a wider green shaded region shown in Fig.~\ref{fig:13} and Fig.~\ref{fig:14}.}
    \label{fig:15}
\end{figure*}

\begin{figure*}
\sidecaption
  \includegraphics[width=12cm]{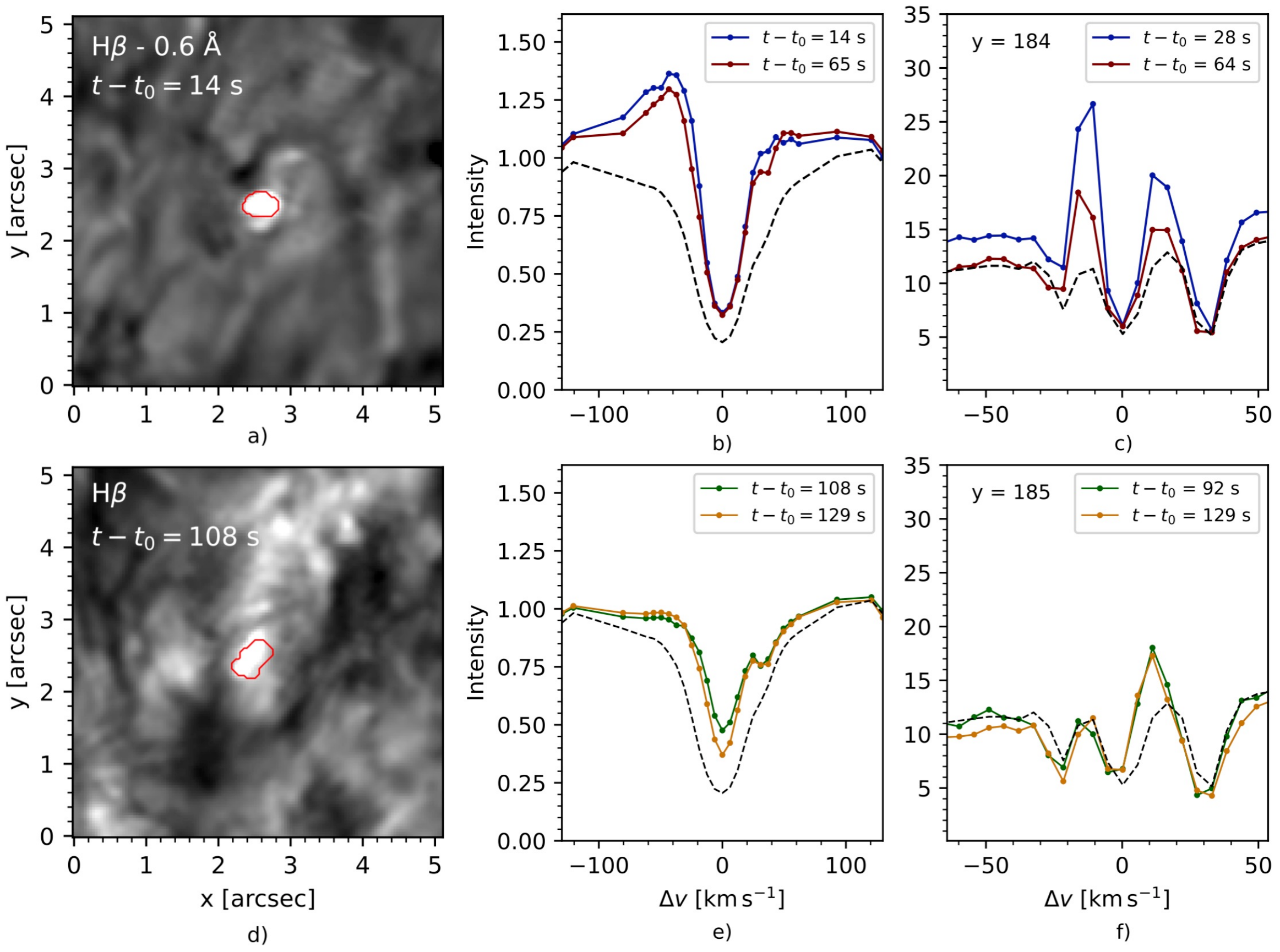}
     \caption{Example of a QSEB with significant emission in the \Mgtriplet\ lines. Panel a) shows the QSEB during the early phase at 14~s which has significant wing enhancement of the \Hbeta\ spectral line. Normalized \Hbeta\ intensity is shown in panel b) at two instances of 14~s and 65~s. Panel c) shows the \Mgtriplet\ profile during this early phase for two timesteps at 28~s and 64~s  from y = 184 in IRIS pixels. Panel d) shows the QSEB at 108~s during the end phase when the line core brightening of \Hbeta\ is more prominent. Spectral profiles for \Hbeta\ are shown in panel e) for timesteps at 108~s and 129~s, and for a location that is located at the top of the QSEB, which is slightly more towards the limb. Similar to panel c), the \Mgtriplet\ profiles are shown in panel f) for two timesteps during the end phase of the QSEB, located at y = 185 in IRIS pixels and have units as DN~s$^{-1}$.}
     \label{fig:16}
\end{figure*}

\section{Results}
\label{sec:results}

\subsection{Temporal and Spatial Analysis of QSEBs and associated UV brightenings}

From the 67 tracked QSEBs and UV brightening pairs, we present 5 events in Fig.~\ref{fig:1}, which shows the full FOV with zoomed-in boxes of the QSEBs and their associated $ 5\sigma $ brightenings.
It is clear from the zoomed-in boxes that the UV brightenings usually occur at some spatial offsets from the QSEB in the direction towards the limb.
Another example of a QSEB - UV brightening pair is shown in Fig.~\ref{fig:2}, which shows the temporal evolution of these events.
The UV brightening occurs almost at the same spatial location as the QSEB. 
The blue cross denoting the center of QSEB lies at the boundary of the UV brightening in many instances, with the UV brightening occurring more towards the direction of the solar limb. 
Even though the UV brightening is detected as a $5\sigma$ event 28~s after the QSEB begins, the same region in SJI is already bright when the QSEB starts. 
This suggests that there could be more UV brightenings associated with QSEBs by keeping a lower threshold on the SJI 1400 images.

The QSEB ends after 57~s, but the UV brightening continues for another 57~s. Further, we look for more such combined QSEB and UV brightening occurrences to gain a better understanding of their temporal and spatial dynamics. 
Within our dataset, we find a total of 67 QSEB - UV brightening event pairs. 
Among these, 59 QSEBs and 59 UV brightenings are uniquely identified, and we note instances where a single UV brightening is associated with multiple QSEBs, typically 2 to 3, suggesting connectivity between them.
Additionally, we observed that sometimes a UV brightening initiates before a QSEB begins, and another UV brightening starts later, close to the QSEB, with some temporal overlap with the QSEB. 

Figure ~\ref{fig:3} presents a detailed visualization of the temporal relationship between QSEBs and UV brightenings. 
The timeline of each of the 67 events is normalized to the duration of each QSEB-UV brightening pair, allowing for a direct comparison across all events. 
From the 67 paired events, 53 coincide temporally to some extent, while the remaining 14, although not overlapping, happen within a span of 35~s of each other’s total duration. 
The possible scenarios of QSEB and UV brightening pairs occurrence are: i) the QSEB and UV brightening start simultaneously, with the UV brightening ending first (e.g., events 1 and 2), ii) the QSEB and UV brightening begin at different times but end at the same time (e.g., events 5 and 29), iii) the QSEB occurs first with some temporal overlap with the UV brightening (e.g., events 33 and 34), iv) the UV brightening occurs first, with some temporal overlap with the QSEB (e.g., events 28 and 32), v) either the QSEB or the UV brightening persists for the entire observed period while the other occurs for some part of the duration (e.g., events 56 and 61) and, vi) either of the QSEB or the UV brightening occurs first, but the other happens with some delay, with no temporal overlap (e.g., events 43 and 45).
The example shown in Fig.~\ref{fig:2} is event number 15. The QSEB started first, and the UV brightening started shortly after 20\%\ of the 114~s combined duration. The QSEB ended at about the middle of the combined duration while the UV brightening reached peak intensity at about 70\% of the combined duration.

Figure~\ref{fig:4} illustrates the spatial distribution of the QSEBs and their corresponding UV brightenings superimposed on a map of the extreme values of $B_\mathrm{LOS}$ within the full FOV. These events tend to cluster in areas where the magnetic field is notably stronger. The map showing these extreme $B_\mathrm{LOS}$ values is also included in the top panel of the figure. The bottom panel shows some QSEBs marked by green circles. These are the events which occur multiple times during the full time series. From the uniquely identified 59 QSEBs, we observed that 24 repeatedly occurred at nine locations shown by the green circles.  
The events in Fig.~\ref{fig:2} with red, green and blue boxes are three of those sites where both QSEBs and their associated UV brightenings occurred more than once during the entire time series. 

The example presented in Fig.~\ref{fig:2} demonstrates that a UV brightening can take place very close to the QSEB. 
The analysis of the 53 co-temporal pairs reveals a broad range of spatial offsets between QSEBs and UV brightenings. 
Figure~\ref{fig:5} offers a statistical overview of these spatial offsets. 
It is important to note that the distances are measured in IRIS pixel units (1 pixel = 0\farcs33), with each histogram bar corresponding to the angular size
of an SST pixel equal to 0\farcs038. 
Panel (a) depicts the distances between the centroids of UV brightenings and QSEBs at each coinciding timestep. 
The majority (71\%) of UV brightenings are situated 2--4 IRIS pixels away from QSEBs. 
A smaller portion (16\%) of events are also found within 4--6 IRIS pixels, while others (13\%) are located closer, within a 2--pixel distance. 
Likewise, panel b) illustrates the separation between UV brightening boundaries and QSEBs, which depicts a slight reduction of the offset ranges.  
Since the boundaries of the UV brightenings are much closer to the QSEBs, we see that the bulk of events (73\%) have boundaries 1--3 pixels away. 
Several are positioned extremely close (14\%), within 1 pixel, while a few events (13\%) are observed 3--5 pixels from the QSEB.

Figure~\ref{fig:6} gives a visual representation of how the UV brightenings occur in the surroundings of their QSEBs. 
The scatter plots vividly map out the spatial distribution of UV brightenings relative to their QSEB counterparts, as if each QSEB were placed at the center of the coordinate system. 
Panel a) presents all co-temporal timesteps for each of the 53 QSEB-UV brightening pairs, utilizing markers with unique combinations of symbols and colors to denote the relative position of the centroids of UV brightenings with respect to their corresponding QSEBs. 
The numerous markers for each event highlight every instance where the QSEB-UV brightening pair occurs together.
We find that markers for individual events are tightly clustered and show minor fluctuations in both distance and orientation relative to the QSEBs over the duration that the QSEB - UV brightening events occur together. 
Approximately 76\% of UV brightenings are located within $\pm 65\deg$ w.r.t the QSEB, with a small spatial offset towards the direction of the solar limb.
In a similar way, panel b) presents the distribution of the boundaries of the UV brightenings with the QSEB at the center. 
Here, the markers used for individual events are the same as for those events in panel a). 
This allows for a clear comparison of the spatial offsets of the centroids and boundaries for each event. 
The event boundaries are significantly nearer to the QSEB, and have a similar spread across the figure as in panel a). 
Similarly, the majority of the events
(75\%) 
reside within the $\pm 65\deg$ vicinity of the QSEB.

\subsection{Analysis of QSEBs captured by the IRIS Slit}
In our dataset, we identified 21 QSEBs which are covered by the IRIS slit. 
For each of these events, we examined the presence of emissions in the \Mgtriplet\ 2798.8~\AA\ and \Siiv\ 1394~\AA\ spectral lines. 
Among these, the spectral profiles of 17 events either exhibited noise or showed enhancements that were not significantly distinguishable. 
However, we present four cases where notable emissions were observed either or both in the core of the \Siiv\ line and in the wings of the two \Mgtriplet\ lines, which are situated between the \MgK\ and \MgH\ lines \citep{2015ApJ...806...14P}. For these cases, the intensity of the spectral lines shown in the figures has been divided by the exposure time of 8~s giving a unit of DN~s$^{-1}$.

\subsubsection{Spectral Analysis of QSEB-1}
In Fig.~\ref{fig:7}, panel (a) presents an example of a QSEB, shown in the blue wing of the \Hbeta\ line. 
Moving to panel (e), the spectra show the characteristic EB profile with clear enhancement in the wings of the \Hbeta\ line, which lasted for 65~s.
Panel (b) displays the line-of-sight magnetic field, $B_\mathrm{LOS}$, and shows that the QSEB originates in a region of negative magnetic polarity near similar magnitude positive polarity.  
In the SJI 1400 observations, we found that the slit was often precisely covering the QSEB location.
For the timesteps when the slit did not sample the region where the QSEB occurred, $5\sigma$ events were also observed in SJI 1400 throughout the QSEB's lifetime.
This can be clearly seen in the accompanying movie.

Figure~\ref{fig:7} presents a snapshot that captures the peak emission in the \Siiv\ 1394~\AA\ line.
This is also the strongest \Siiv\ emission observed for co-temporal, co-spatial QSEB, and UV brightenings detected in this work.
The \Siiv\ profile exhibits a redshift and has a Doppler shift of $+18 \mathrm{~km} \mathrm{~s}^{-1}$.
Note that the panels f), g) and h) show the spectral profiles for \MgK, \Mgtriplet, and \Siiv, which are offset by 1 IRIS pixel toward the limb from the QSEB location.

The \MgK\ profile in panel f) is broader than the quiet reference profile but does not stand out as different from profiles in the vicinity. 
This is also valid for the \Mgtriplet\ lines in panel g) which does not show any enhancement. 
However, the associated movie displays occasional intensity enhancements in the wings of the \Mgtriplet\ line throughout the event's duration. 
Additionally, the \Siiv\ 1394~\AA\ line shows increased emission and subtle broadening at several timesteps concurrent with the QSEB. 
These panels further highlight regions shaded in pink for the \Mgtriplet\ and green for the \Siiv\ line, which we use 
for generating a time sequence of spectroheliograms for these wavelength ranges in Fig.~\ref{fig:8} and Fig.~\ref{fig:9}. 
These show the evolution of the integrated line intensities over time and space ($x$-axis) against the pixels along the slit ($y$-axis). 
Specifically, in Fig.~\ref{fig:8}, the spectroheliogram reveals that the intensity increase in the \Mgtriplet\ starts approximately half a minute before the start of the QSEB and at the location where the QSEB occurs and continues while the QSEB is in progress. The location and duration of the QSEB in \Hbeta\ is marked with solid lines.
Figure~\ref{fig:9} presents the spectroheliogram for \ion{Si}{iv} 1394~\AA, highlighting that the peak emission for \ion{Si}{iv} occurs between 08:40:24 to 08:41:08 UT shown by the yellow pixels. 
This also corresponds to the time when the QSEB exhibits its maximum wing enhancement of \Hbeta\ line and happens in a region 1--2 IRIS pixels above the location where the QSEB occurs.

\subsubsection{Spectral Analysis of QSEB-2}
Figure~\ref{fig:10} illustrates another QSEB, visible in the \Hbeta\ blue wing. 
Panel e) shows the spectral profile, which exhibits notable, although moderate, wing enhancement in the \Hbeta\ line, observable for about 3~min. 
Panel (b) shows the ($B_\mathrm{LOS}$) magnetic field, with the QSEB emerging in a negative polarity patch amidst nearby positive polarity regions.
In SJI 1400, we see that throughout the event's duration, the slit frequently captures the QSEB directly. 
However, when the QSEB occurs outside the immediate area of the slit, the nearby areas are marked by distinct high-intensity events that exceed the $5\sigma$ level in several timesteps. 
Figure~\ref{fig:10} presents the images and spectral profiles corresponding to the moment when the emission of the \Siiv\  1394~\AA\ line reaches its maximum intensity for this case. 
Here, the profiles are also from the location at a spatial offset of 1 IRIS pixel towards the limb.
The \Siiv\ profile is slightly blue-shifted with a Doppler shift of  $-7.3 \mathrm{~km} \mathrm{~s}^{-1}$.
The \MgK\ line and the \Mgtriplet\ profiles exhibit no notable enhancements for this particular snapshot.
But, the accompanying movie highlights a consistent enhancement in the \Mgtriplet\ wings, persisting throughout the entire duration of the event, for example, at time 08:53:08 UT. 
Moreover, the \ion{Si}{iv} 1394~\AA\ line shows both increased emission and a subtle broadening in sync with the QSEB. 
The \ion{Mg}{ii}~triplet spectroheliogram (Fig.~\ref{fig:11}) shows that the most intense (brightest yellow) part is located just one pixel above the QSEB site. 
This intensity begins to surge at the start of the QSEB, reaching its maximum between 08:53:30 and 08:54:06 UT. 
Interestingly, the strongest emission in the wings of the \Hbeta\ line of the QSEB coincides with this phase of increased intensity, occurring around 08:53:43 UT. 
Figure~\ref{fig:12} for \ion{Si}{iv} 1394~\AA\ highlights the peak emission timing, from 08:53:12 to 08:54:07 UT, occurring one pixel above the QSEB's location and aligning with the \Hbeta\ line's maximum wing enhancement.
Another subtle increase in the intensity of \Hbeta\ wings occurs towards the end of the QSEB, for instance, at 08:54:48 UT, which is also accompanied by an enhancement in the \Siiv\ 1394~\AA\  line intensity. This is visible as light green patches at an offset of 1 IRIS pixel in the spectroheliogram of Fig.~\ref{fig:12}, just before the QSEB ends.

\subsubsection{Spectral Analysis of QSEB-3}

The third example, captured in Fig.~\ref{fig:13}, illustrates the scenario where the UV brightening occurs before the QSEB.
From the associated movie with this figure, it can be seen that the UV brightening initiates before the QSEB, starting at 08:21:55 UT, with its intensity peaking at 08:23:11 UT. The \Siiv\ 1394~\AA\ line has predominantly a red-shifted profile during the entire time. 
Figure~\ref{fig:13} shows the images and spectral profiles for the instance with maximum emission and significant broadening of the \Siiv\ 1394~\AA\ line which has a Doppler shift of $+54.2 \mathrm{~km} \mathrm{~s}^{-1}$ at this time.
As can be seen from panels (e) and (g), the \Hbeta\ and \Mgtriplet\ profiles are almost similar to the average quiet Sun profile shown by the dashed lines. 
Interestingly, during periods of strong \Siiv\ enhancement, there is a small reduction in the intensity of the \ion{Mg}{ii} k$_\mathrm{2r}$ spectral feature, falling below the quiet Sun average. 
The \MgK\ core is red-shifted as shown in Fig.~\ref{fig:13}, and there is a decrease in intensity of k$_\mathrm{2r}$, suggesting that the line core opacity is shifted to that of k$_\mathrm{2r}$. 
This creates an opacity window that increases the intensity of k$_\mathrm{2v}$ of \MgK\, indicating downflows in the region. 
Similar opacity shifts and opacity window effects have been discussed by \cite{2019A&A...631L...5B} for the formation of \MgK\ spectral line in type-II spicules.

The QSEB itself begins at 08:25:24 UT and concludes at 08:26:49 UT, with its peak wing enhancement occurring at 08:26:06 UT. 
Figure~\ref{fig:14} corresponds to the snapshot when the QSEB has maximum \Hbeta\ wing emission.
Following this peak, although the wing enhancement diminishes, the line core brightening becomes more pronounced. 
After examining the \Hbeta\ line core, it was noted that the onset of brightening occurred at 08:26:06 UT (time of the snapshot in Fig.~\ref{fig:14}) and progressed towards the top of the QSEB in the direction of the solar limb. 
At the same time, there was a subtle emission in the wings of the \ion{Mg}{ii} triplet line too, which is shown at the location of the QSEB. 
The \Siiv\ emission continues until approximately 08:26:21 UT. 
The QSEB emerges over a region of positive magnetic polarity as seen in panel (b) of Fig.~\ref{fig:14}, closely surrounded by areas of negative polarity. 
At instances when the slit drifts from the immediate area, the region's activity remains detectable through $5 \sigma$ brightening events, although only in a couple of timesteps. 
Panel (h) in both figures shows a wider green-shaded region compared to the previous cases due to the broadening and redshift of the \Siiv\ profiles. 
The spectroheliogram for this shaded region is shown in Fig.~\ref{fig:15}, which distinctly illustrates that the \Siiv\ emission occurs at the same location as the QSEB. The emission initiates prior to the QSEB and continues with lesser intensity during the QSEB.

\subsubsection{QSEB with the strongest emission in the \Mgtriplet\ lines}
The QSEB shown in Fig.~\ref{fig:16} occurs for about 2~min 23~s and is characterized initially by significant enhancements in the \Hbeta\ line wings. 
At 14~s from the start of the QSEB, the enhancement of the wings reaches its maximum, subsequently followed by the most pronounced emission in the \Mgtriplet\ lines wings. This occurs around 28~s into the event and is located at an offset of 1 IRIS pixel (at y = 184). The QSEB occurs at y = 183 in IRIS pixels. 
An increase in the wings' intensity of the \Mgtriplet\ lines is also noted at 64~s at the same location as before, coinciding with an intensity enhancement in the wings of the \Hbeta\ line at 65~s. 
The QSEB initiates with a notable brightening of the \Hbeta\ line core, which, after 86~s, intensifies further, accompanied by a decrease in wing intensity. 
A snapshot of the QSEB at a later phase at 108~s in Fig.~\ref{fig:16} shows the instance with maximum line core brightening of the \Hbeta\ line. 
It is also observed that the line core brightening occurs more at the top of the QSEB, which is towards the solar limb, compared to the profiles at 14~s and 65~s.
Additionally, emissions from the \Mgtriplet\ lines wings are observed at 92~s and 129~s, although less prominent than the peak emission at 28~s. This emission has also shifted by 1 IRIS pixel towards the limb, compared to the spectral profiles at 28~s and 64~s.
\section{Discussion}
\label{sec:discussion}
We have studied the transition region response to QSEBs, with a particular emphasis on the spatial and temporal interplay between QSEBs and UV brightenings. This was achieved by detecting QSEBs in \Hbeta\ and UV brightenings from the IRIS imaging and spectrograph data. We conclude that QSEBs can show significant emissions in the SJI 1400 and also in the \Siiv\ 1394~\AA\ and \ion{Mg}{ii} 2798.8~\AA\ triplet spectral lines.

We implemented the \textit{k}-means clustering algorithm and detected 1423 QSEBs in a quiet Sun region near the northern solar limb, across a 51~min observation period.
This is a lower detection rate than \cite{2022A&A...664A..72J} who identified 2809 QSEBs during a one hour \Hbeta\ time series and \cite{2024A&A...683A.190R} who detected 961 \Hbeta\ QSEBs in a 24~min time series. 
We attribute their higher detection rates mostly to their better seeing quality data. Both these studies show there is a strong correlation between seeing quality and number of QSEB detections. 

In order to clearly see the temporal evolution in the SJI 1400 data, we needed to restrict our analysis to events that lived for at least three time steps. 
Given the low signal level in quiet Sun, we settled for 8~s exposure time for the IRIS observing program.
This effectively resulted in 18~s cadence for the SJI data. 
We focused on long-lived QSEBs, which yielded 453 QSEBs that lasted for at least one minute. 
The UV brightenings were detected using a threshold of 5$\sigma$ above the noise level to detect significant \Siiv\ emissions. 
The threshold of $5\sigma$ is conservative, and there could be more UV brightenings associated with QSEBs by keeping a lower threshold on the SJI 1400 images. From the analysis of QSEBs with the IRIS slit on top, we found that the events which show some emission in the \Siiv\ 1394~\AA\ also have nearby regions in SJI 1400 which are detected through the threshold of 5$\sigma$. Hence, this threshold gives us the strongest events from the SJI 1400 for studying their evolution with the QSEBs.
This approach led to the identification of 1978 UV brightening events. 
To further explore the spatial and temporal proximity between QSEBs and UV brightenings, we searched for UV brightenings occurring within a 4\arcsec $\times$ 4\arcsec\ area surrounding the QSEBs and within 35~s of QSEBs duration.  
This resulted in the identification of 67 QSEB-UV brightening pairs for detailed analysis. 
From the total number of QSEBs identified, 21 were directly in the field of view of the IRIS slit.
Among these, 4 (20\%) were distinguished by their enhanced emissions in either or both of the \Mgtriplet\ 2798.8~\AA\ and \Siiv\ 1394~\AA\ spectral lines. 
Within the 67 QSEB-UV brightening pairs, we recognized 59 unique QSEBs and 59 unique UV brightenings and observed cases where a single UV brightening is located near more than one QSEB at the same time. 
A similar example of two EBs near a UV burst in an active region was shown by \cite{2019ApJ...875L..30C}.

Magnetic reconnection in the chromosphere is a complex process. For example, \cite{2019A&A...626A..33H} found an EB and a UV burst along an extended current sheet where reconnection occurs first at heights of the upper photosphere to the lower chromosphere and later in the upper chromosphere. In other scenarios, we know that magnetic reconnection occurs in the higher layers of the solar atmosphere, such as during large-scale eruptive events like solar flares. The process of reconnection in the partly ionized chromosphere is less understood \citep[see, e.g.,][]{2013PhPl...20f1202L, 2014SSRv..184..107L, 2018SSRv..214...58B, 2020A&A...633A..66N} and also requires further investigation to elucidate the role of particle acceleration \citep[see, e.g.,][]{1998A&A...337..294H}.

We observed that a UV brightening might coexist with a QSEB, with a subsequent UV brightening appearing near the same QSEB at a later time.
Of the 67 pairs analyzed, 53 demonstrate co-temporal existence, with either the QSEB or UV brightening starting or ending first. 
The 14 pairs without direct temporal overlap appear nearly half a minute apart, suggesting that these pairs of QSEBs and UV brightenings could also be related to each other. 
The analysis indicates a higher likelihood of QSEBs beginning first, with 38 (57\%) occurrences of QSEBs starting before UV brightenings. 
This suggests the onset of magnetic reconnection in the lower atmosphere, which then propagates to higher layers. 
In our analysis of raster data for QSEBs, we find two types of scenarios. 
For the first two cases of QSEBs, there is a noticeable enhancement in the  \Siiv\ 1394~\AA\ spectral line coinciding with the QSEB events, but beginning after the start of the QSEB.
This enhancement is also synchronous with the peak of \Hbeta\ wing enhancement for the QSEBs.
The third QSEB, captured by the IRIS slit, presents the second scenario with pronounced \Siiv\ 1394~\AA\ spectral line emissions at the location of the QSEB. 
These emissions precede the initiation of the QSEB but also continue during its occurrence. 
The second scenario is corroborated by 24 events (36\%) where UV brightenings start before QSEBs, suggesting a top to bottom propagation of magnetic reconnection from the upper to the lower atmosphere.
In the case of the 5 QSEB-UV brightening pairs that initiate simultaneously, the UV brightenings tend to end first, with QSEBs exhibiting a longer duration.

For the 53 co-temporal pairs, we found that the majority (84\%) of the UV brightenings occur within 4 IRIS pixels ($\approx$~1000~km) of the QSEBs and usually occur in the direction of the solar limb. 
The \textit{k}-means clustering algorithm was also applied on a different dataset of a quiet Sun region, observed on 25 July 2021. 
By checking the evolution of UV brightenings with the QSEBs, it was noted that the UV brightenings of this region also show a spatial offset and tend to occur towards the limb direction.
For the QSEBs sampled by the IRIS slit, the most significant emissions in both the \Mgtriplet\ 2798.8~\AA\ and \Siiv\ 1394~\AA\ lines occur at spatial offsets of 1--2 IRIS pixels (244 -- 487 km) in the direction of the solar limb.
In the instances of two QSEBs documented by \cite{2017ApJ...845...16N} that are accompanied by brief increases in intensity in the SJI 1400, no spatial offset is detected despite their observation region being closer to the limb. 
\cite{2015ApJ...812...11V} provides observational support for spatial offsets between EBs and UV bursts, suggesting that the tops of EBs exhibit higher temperatures. 
Additionally, \cite{2019ApJ...875L..30C} report spatial offsets observed between 20 EBs and UV bursts, with UV bursts occurring at the top of the EBs. 
They estimate that the height difference between these events could range a few hundred km.
Furthermore, \cite{2019A&A...626A..33H} explain through their simulation that EBs and UV bursts can occur co-spatially and co-temporally, with magnetic reconnection happening at various heights of an extended current sheet. 
The spatial offsets in these events could be a consequence of how the current sheet is oriented.
Another possibility is that these spatial offsets are due to projection effects because of the observation region being close to the solar limb.

Examination of the 59 unique QSEBs revealed that 24 of them repeatedly appeared at nine particular sites within the FOV in areas that are marked by concentrations of stronger magnetic fields. 
This behaviour of recurring activity at the same location is also observed for EBs and UV bursts \citep{2015ApJ...812...11V}.

Across all three instances featuring \Siiv\ 1394~\AA\ emissions in this work, we noted subtle broadening of the line profile and Doppler shifts.
However, none of these instances displayed characteristics typical of UV bursts in the \Siiv\ 1394~\AA\ line, which are defined by strong emission (peak at hundreds to thousands times the surrounding level) and broad wings that show the absorption features of \ion{Ni}{ii} and \ion{Fe}{ii} superimposed on this line.
A similar analysis of the 25 July 2021 quiet Sun region revealed even weaker \Siiv\ emissions for the QSEBs, which were sampled by the IRIS slit.
Most QSEBs, when observed directly under the IRIS slit, show \Siiv\ 1394~\AA\ spectral profiles of similar magnitude as the quiet Sun's average profile.
The cases highlighted in this study represent those with the most significant enhancements in the \Siiv\ line profile we find in our data. 
Emission in the \Siiv\ line can signify heating to temperatures of 80,000~K under equilibrium conditions \citep[see, e.g.,][]{2014Sci...346C.315P}.
\cite{2019A&A...626A..33H} found that the current sheet heats from 10,000~K to 1~MK within 20~s from 700 km up to 3500 km where UV burst occurs in their simulation.
\cite{2017ApJ...845...16N} reported a QSEB sampled by the IRIS slit with enhanced emissions in the \Siiv\ 1394~\AA\ line profiles, which have intensity higher than we find in our data.

Among the slit events, the first two QSEBs show a wider than average \MgK\ line core emission feature, 
along with emission in the wings of the \Mgtriplet. 
This kind of behaviour of \ion{Mg}{ii} lines was studied by \cite{2015ApJ...806...14P} using both simulations and observations.
In such a scenario, the heating takes place deeper in the atmosphere close to the temperature minimum, with overlying cool material as indicated by absorption in the core of \Mgtriplet\ with emission limited to its wings. 
The broader \MgK\ profiles suggest that they form much deeper than usual.
\cite{2016A&A...593A..32G} 
provide evidence that strong brightenings observed in \ion{Mg}{ii} lines, form within a broad altitude range, from 400 to 750 km. Some of the events analysed by them also have signatures in the \Siiv\ 1394~\AA\ spectral line and are examples of UV bursts described by \cite{2014Sci...346C.315P}.
The first two QSEBs of our analysis are also accompanied by the intensity increase, broadening, and small Doppler shifts of the \Siiv\ line. 
Simultaneous emissions in these spectral lines have been reported by \cite{2015ApJ...812...11V} for the case of EBs and UV bursts. 

The third and fourth QSEBs of the slit events display unique behaviours between the \Mgtriplet\ lines and the \Hbeta\ line. 
For the third QSEB, a period of peak \Hbeta\ wing emission coincides with subtle emissions in the wings of the \Mgtriplet\ line at the location of the QSEB.
The fourth QSEB shows the strongest \Mgtriplet\ emissions in this work (at 28~s and 64~s as shown in Fig.~\ref{fig:16}) and at an offset of 1 IRIS pixel ($\approx$ 244 km) towards the limb from the QSEB, which notably occur after strong wing enhancements of \Hbeta\ line, and indicate a correlated atmospheric response. 
The magnitude of the emission at 28~s is also comparable to the case of the quiet Sun \Mgtriplet\ profile reported by \cite{2015ApJ...806...14P}. The example of QSEB sampled by the IRIS slit, presented by \cite{2017ApJ...845...16N} also shows similar enhancement in the wings of the \Mgtriplet\ lines.
When the fourth QSEB moves slightly towards the limb and manifests as \Hbeta line core brightening during the end phase, the enhancement in the \Mgtriplet\ is also shifted by 1 more pixel towards the limb. 
This spatial and temporal interplay between the enhancement in the \Hbeta\ line and emissions in the \Mgtriplet\ line suggests upward propagating magnetic reconnection along the vertically elongated current sheets from the photosphere to the lower chromosphere. 
The shifting of line core brightening of the \Hbeta\ line towards the limb, compared to the wings, is a common phenomenon observed in 93\% of the cases of \cite{2022A&A...664A..72J}. 
\cite{2024A&A...683A.190R} further noted the sequential appearance of QSEBs in \Hbeta\ and then in H$\varepsilon$, implying an upward movement of magnetic reconnection in the atmosphere.

We regard our method of associating QSEBs to UV brightenings as conservative: we have avoided ambiguous associations and did not consider \Siiv\ emission related to other phenomena in connection with magnetic bright points or nearby spicules. 
Out of the 453 long-lived QSEBs, we find 67 (15\%) QSEB-UV brightening pairs.
It is likely that there are more QSEBs with associated UV brightenings but with shorter lifetimes than the ones we analyze here. Our analysis was limited to longer duration events due to the difference in cadence between SST and IRIS.
To study the UV brightenings related to short-lived QSEBs, UV observations with faster cadence are required for comparing the evolution of these events.
In the study of cancellation of internetwork magnetic fields in quiet Sun, \cite{2018ApJ...857...48G} reported that approximately 76\% of the events exhibited co-spatial bright grains in the SJI 1400 observations. 
We speculate that a subset of these events that occur due to magnetic reconnection could likely be QSEBs, as they are characterized by similar brightenings in the SJI 1400 as those detailed in our examples.
\cite{2024A&A...683A.190R} suggest that there may be at least 750,000 QSEBs occurring at the Sun at any time. 
Our study indicates that 
a subset of these may 
be associated with heating leading to visibility in transition region diagnostics.
The magnitude of the emissions in the chromospheric and transition region diagnostics for the quiet Sun is also lower than in the case of EBs in active regions. 
While some of the long-lived QSEBs in this study result in localized heating up to transition region temperatures, they are a small fraction of the total QSEB population. This suggests that the long-lived QSEBs do not contribute significantly to the global heating of the upper atmosphere.

\begin{acknowledgements}
The Swedish 1-m Solar Telescope is operated on the island of La Palma
by the Institute for Solar Physics of Stockholm University in the
Spanish Observatorio del Roque de los Muchachos of the Instituto de
Astrof{\'\i}sica de Canarias.
The Institute for Solar Physics is supported by a grant for research infrastructures of national importance from the Swedish Research Council (registration number 2017-00625).
This research is supported by the Research Council of Norway, project number 325491, 
and through its Centres of Excellence scheme, project number 262622. 
IRIS is a NASA small explorer mission developed and operated by LMSAL with mission operations executed at NASA Ames Research center and major contributions to downlink communications funded by ESA and the Norwegian Space Agency.
J.J. is grateful for travel support under the International Rosseland Visitor Programme. 
We made much use of NASA's Astrophysics Data System Bibliographic Services.
\end{acknowledgements}

\bibliographystyle{aa}
\bibliography{bhatnagar_ref} 

\end{document}